\def\bar{\begin{eqnarray}}
\def\ear{\end{eqnarray}}
\def\bb{\bibitem}
\def\eqi{\begin{equation}}
\def\eqf{\end{equation}}
\def\eqia{\begin{eqnarray}}
\def\eqfa{\end{eqnarray}}

\def\oc2{$\mathcal{O}(c^{-2})$}

\documentclass[11pt]{article}
\usepackage{amsmath,amsthm,amscd,amssymb}
\usepackage{latexsym}
\usepackage{graphicx,epsfig}

\begin{document}

\noindent{\bf \LARGE{Reply to ``On the Systematic Errors in the
Detection of the Lense-Thirring Effect with a Mars Orbiter '', by
Giampiero Sindoni, Claudio Paris and Paolo Ialongo }}
\\
\\
\\
{Lorenzo Iorio}\\
{\it Viale Unit$\grave{a}$ di Italia 68, 70125\\Bari (BA), Italy
\\tel. 0039 328 6128815
\\e-mail: lorenzo.iorio@libero.it}

\begin{abstract}
In this note we reply to the criticisms by Sindoni, Paris and
Ialongo concerning some aspects of the recent frame-dragging test
performed by Iorio with the Mars Global Surveyor (MGS) spacecraft
in the gravitational field of Mars.
\end{abstract}



\section{Introduction}
The remarks by Sindoni et al (2007) to the analysis by Iorio
(2007a)  mainly concern with an alleged huge underestimation of
the impact of various sources of systematic errors, both of
gravitational and non-gravitational origin, in the MGS orbital motion.

%
\section{The gravitational perturbations}
In regard to the impact of the mismodelling in the even zonal
harmonic coefficients  of the multipolar expansion of the Martian
gravitational field on the out-of-plane portion of the MGS trajectory, a mere
formal evaluation can be done following, e.g., (Rosborough and
Tapley 1987; Iorio 2003). By using the MGS95J solution (Konopliv
et al. 2006) and truncating the calculation at degree $\ell=20$
for the sake of simplicity, we get an average bias of $49.297$ m
over the same time span of the analysis by Iorio (2007a), i.e.
$30$ times larger than  the predicted Lense-Thirring shift; the
inclusion of the other higher degree terms would certainly make it
bigger. In fact, the average of the MGS out-of-plane RMS orbit overlap
differences by (Konopliv et al. 2006) over the observational time span used by Iorio
(2007a) amounts to 1.613 m only; there is no trace at all of the
very huge bias postulated by Sindoni et al (2007) whose existence
is, thus, neatly and unambiguously ruled out by the data processed by Konopliv et al. (2006).
\section{The non-gravitational perturbations}
In regard to the non-conservative forces, Sindoni et al. (2007)
yield a total un-modelled  non-gravitational acceleration of
$\approx 10^{-11}$ m s$^{-2}$ which is of the same order of magnitude
of the Lense-Thirring acceleration induced by Mars on MGS. They do
not present detailed calculation of the effect of such an
acceleration on the normal portion of the MGS orbit, but some
simple considerations can be traced: a perturbing out-of-plane
force 6.7 times larger than the Lense-Thirring one, as claimed by
Sindoni et al (2007), and having the same time signature should
induce a 10.8 m cross-track shift, on average, over the considered
time span. Again, such a bias is neatly absent from the data.
Time-dependent, long-period, i.e. averaged over one orbital revolution which covers about 2 hr, 
signatures would, instead, be averaged
out. It is just the case, as shown by the time-varying patterns of the main non-gravitational accelerations 
over 12 hr  presented in (Lemoine et al. 2001).
Moreover, as clearly stated in (Konopliv et al. 2006) and (Lemoine et al. 2001),
the along-track and radial portions of the MGS orbit$-$left unaffected
by the Lense-Thirring force$-$ are primarily perturbed by the
non-gravitational forces: indeed, the along-track empirical
accelerations fitted by Konopliv et al. (2006) amount just to
$\approx 10^{-11}$ m s$^{-2}$.
\section{Conclusions}
In conclusion, we cannot find evidence of the total 10
km corrupting shift postulated by Sindoni et al. (2007) in the time series of the
out-of-plane RMS orbit overlap difference of MGS. It must be recalled that 
RMS of orbit solution overlaps are commonly used in satellite
  geodesy as useful and significant indicators of the
  overall orbit accuracy (Tapley et al. 2004), i.e., 
they, among other things, account for all the mis-modelled/un-modelled forces acting on the 
spacecraft, independently of their physical origin: they concisely tell us the whole about any sort of
errors (systematic, measurement, modelling, etc.). Any serious speculations 
cannot leave aside this basilar, but fundamental, fact.


\end{document}